\documentclass[article,onecolumn,superscriptaddress,nofootinbib,10pt]{revtex4}
\usepackage{graphicx}
\usepackage{amssymb}
\usepackage{epstopdf}
\usepackage{amsmath}
\usepackage{hyperref}
\usepackage{mathrsfs}
\usepackage{varioref}
\usepackage{tikz}
\usepackage{lmodern}
\usetikzlibrary{decorations.pathmorphing}
\tikzset{snake it/.style={decorate, decoration=snake}}
\usepackage{fancybox}
\expandafter\ifx\csname package@font\endcsname\relax\else
\expandafter\expandafter
\expandafter\usepackage
\expandafter\expandafter
\expandafter{\csname package@font\endcsname}%
\fi
\hypersetup{
    bookmarks=false,         
    pdfstartview={FitH},    
    colorlinks=true,       
    linkcolor=blue,          
    citecolor=blue,        
    filecolor=blue,      
    urlcolor=blue           
}
\DeclareFontFamily{OT1}{pzc}{}
\DeclareFontShape{OT1}{pzc}{m}{it}%
             {<-> s * [0.900] pzcmi7t}{}
\DeclareMathAlphabet{\mathscr}{OT1}{pzc}%
                                 {m}{it}
\labelformat{section}{Section #1} 
\labelformat{subsection}{Section #1} 
\labelformat{equation}{Eq.(#1)} 
\labelformat{figure}{Fig.~#1} 
\labelformat{subfigure}{Fig.~\thefigure#1} 
\labelformat{table}{Tab.~#1} 
\newcommand{\be}{\begin{equation}}
\newcommand{\ee}{\end{equation}}
\newcommand{\bea}{\begin{eqnarray}}
\newcommand{\eea}{\end{eqnarray}}

\begin{document}

\title{Unruh effect for inertial observers through vacuum correlations}

\author{Kinjalk Lochan,}%
\email{kinjalk.lochan@gmail.com}%
\affiliation{Department of Physical Sciences, IISER Mohali, Manauli 140306, India}
\affiliation{IUCAA, Post Bag 4, Ganeshkhind, Pune University Campus, Pune 411 007, India}
\author{Sumanta Chakraborty}%
\email{sumantac.physics@gmail.com}%
\affiliation{Department of Theoretical Physics, Indian Association for the Cultivation of Science, Kolkata 700032, India}
\affiliation{IUCAA, Post Bag 4, Ganeshkhind, Pune University Campus, Pune 411 007, India}
\author{T. Padmanabhan}%
\email{paddy@iucaa.in}
\affiliation{IUCAA, Post Bag 4, Ganeshkhind, Pune University Campus, Pune 411 007, India}

\begin{abstract}
We study a dynamic version of the Unruh effect in a two dimensional collapse model forming a black hole. In this two-dimensional collapse model a scalar field coupled to the dilaton gravity, moving leftwards, collapses to form a black hole. There are two sets of asymptotic ($t\to\infty$) observers, around $x\to\infty$ and $x\to-\infty$. The observers at the right null infinity witness a thermal flux of radiation associated with time dependent geometry leading to a black hole formation and its subsequent Hawking evaporation, in an expected manner. We show that even the  observers at left null infinity find themselves in thermal ambiance, without experiencing any change of spacetime geometry all along their trajectories. They remain as inertial observers in a \emph{flat} region of spacetime  where curvature tensor identically vanishes in a portion of full spacetime.  These observers find the state of the quantum field in a late time thermal configuration, with \emph{exactly the same} temperature as measured by 
the observers at right null infinity, {\it despite being inertial in flat spacetime region throughout their history}. This is very closely related to the standard Unruh effect in the flat spacetime, except for a key difference --- since they are inertial throughout and {\it have no causally connected source in the past light cone to account for what they see}. The result arises from quantum correlations which extend outside the past light cone and is conceptually similar to the EPR correlations.
\end{abstract}
\maketitle
\section{Introduction} 

Hawking radiation from a black hole \cite{Hawking:1974sw,Hawking:1974rv,Birrell:1982ix,Helfer:2003va,Fabbri:2005mw,
Mukhanov-Winitzky,Parker:2009uva,gravitation, Visser:2001kq,Takagi:1986kn} and the Unruh radiation \cite{Davies:1974th,Unruh:1976db,gravitation,Birrell:1982ix,Parker:2009uva,Mukhanov-Winitzky} in the Rindler frame have very similar mathematical properties. In the context of an eternal black hole, the Hartle-Hawking vacuum state of a quantum field will appear as thermal for static observers in the right wedge. This arises because the modes of the quantum field in the region inaccessible to the observer are traced out. Similarly, a uniformly accelerated observer in the right wedge of the flat spacetime will associate a thermal nature to the global, inertial, vacuum state, again obtained by tracing out the modes inaccessible to her (on the left wedge) \footnote{By tracing out we mean that the relevant mode function will have only a  compact support on the Cauchy surface and the behaviour outside the region of support would not contribute in integrations to be performed.}. Both situations are time-reversal 
invariant; while the relevant observers see an ambient thermal radiation, they do not associate a flux of particles with this radiation. In the rest of the analysis, by the word ``radiation'' we will imply scenarios where the wave modes are thermally populated {\emph{but} there may not be any physical energy momentum tensor (and hence flux) associated with it.}

A somewhat different situation arises in the case of a black hole formed by collapsing matter, with the quantum state being the Unruh vacuum (or, the In-vacuum) at very early times. In this case, at late times, observers far away from the collapsing body detect a flux of particles with a spectrum which is thermally populated. The energy carried away by the particles is ultimately obtained from the mass of the collapsing body and this leads to the concept of black hole evaporation. The mathematical description of this process takes into account; (i) the change in geometry due to the collapse process and (ii) the formation of event horizon leading to inaccessibility of a region from future asymptotic observers. With future applications in mind, let us briefly recall the key concepts.
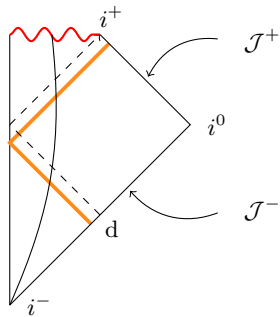
\begin{figure}[h] 
    \begin{center}
        \begin{tikzpicture}[scale=0.8]
        
            \draw[-] (0,-3) -- (0,1.5);
            \draw[dashed,->] (0,0) -- (1.5,1.5);
            \draw[orange!90,line width=0.5mm] (1.65, 1.35) -- (0,-0.3) -- (1.35,-1.65);
       
            \draw[red,thick, snake it] (0,1.5) -- (1.5,1.5);

            \draw[black] (0,-3) -- (3,0) -- (1.5,1.5);
            \draw[black, dashed] (1.5,-1.5) -- (0,0);
        
            \draw[black,bend left=15] (0.7,1.5) to (0,-3);

            \node[label=right:$i^+$] (1) at (1.2,1.8) {};
            \node[label=right:$i^-$] (2) at (0,-3) {};
            \node[label=right:$i^0$] (3) at (3.0,0) {};
            \node[label=below: d](4) at (1.7,-1.3) {};
            
            \node[label=right:$\mathcal{J}^+$] (5) at (3.6,1.4) {};
            \node[label=right:$\mathcal{J}^-$] (6) at (3.6,-1.4) {};

            \draw[->,bend right=35] (5) to (2.3,0.8);
            \draw[->,bend left=35] (6) to (2.0,-1.1);
           
        \end{tikzpicture}
    \end{center}
    \caption{Penrose Diagram For Schwarzschild collapse}
    \label{fig:penrose00}
\end{figure}
In standard 3+1 dimensional collapse, in which classical matter collapses to form a black hole,  an apparent horizon is formed, which grows and --- at a  certain stage ---  an event horizon is formed thereby producing a black hole region. The last null ray originating from the past null infinity $\mathcal{J}^{-}$  at the null co-ordinate $u=d$, in the double null co-ordinate system \cite{Hawking:1974sw,gravitation,Birrell:1982ix,Mukhanov-Winitzky,Parker:2009uva}, reaching future null infinity $\mathcal{J}^{+}$ defines the location of the event horizon. Note that we have taken $u$ as the advanced time coordinate, which is different from the standard choices, where usually $v$ is referred to as advanced coordinates. We will follow this convention throughout. An asymptotic observer on $\mathcal{J}^{+}$ has no causal connection with events inside the event horizon. Further, whole of  $\mathcal{J}^{+}$  derives its complete past causal support (e.g., the thick 
orange line in \ref{fig:penrose00}) from only a \textit{part of} $\mathcal{J}^{-}$, which lies prior to the ray forming event horizon \footnote{Since the rays suffering gravitational lensing do not undergo any frequency shift, they are typically discarded in the standard Hawking effect and we will also not be concerned about them in this paper hereafter.}. 

A wave mode which originates from $\mathcal{J}^{-}$ and reaches $\mathcal{J}^{+}$ also experiences a change in background geometry in the process. In order to obtain the Bogoliubov coefficients between the asymptotic observers \footnote{We can think of observers moving with large speeds and reaching future timelike infinity. These observers, in an approximating limit, may be mimicked through late time observers on $\mathcal{J}^{+}$.} on $\mathcal{J}^{-}$ and $\mathcal{J}^{+}$, we need to re-express a set of complete mode functions $v_{\omega}$ suited to observers on $\mathcal{J}^{+}$ in terms of complete set of mode functions $u_{\omega}$, defined equivalently on $\mathcal{J}^{-}$. When we trace back the out-going modes $v_{\omega}$ on $\mathcal{J}^{+}$ through the center, (i.e., through $r=0$ with $r$ being the Schwarzschild radial co-ordinate) onto $\mathcal{J}^{-}$, we  see that the mode functions suited to $\mathcal{J}^{+}$ have support only on the portion of mode functions $u_{\omega}$ on $\mathcal{J}^{-
}$. 

Therefore the Bogoliubov coefficients are obtained by taking inner products of $u_{\omega}$ and $v_{\omega}$ on a portion of $\mathcal{J}^{-}$, i.e., below the line $u=d$ \cite{Hawking:1974sw,Birrell:1982ix,Parker:2009uva,gravitation}. (This has an effect similar to that of partial tracing out of modes in the case of eternal black hole.) Furthermore, we now also have to take into account the change of geometry experienced by  the mode function $u_{\omega}$  when we trace it back to the past null infinity. However, if we are interested only in the late time behavior of $u_{\omega}$, we can use ray-optics approximation \cite{Hawking:1974sw,Birrell:1982ix,Parker:2009uva,gravitation} for tracing back the mode functions. {\it Thus, the late time behavior of the out going asymptotic future modes is essentially controlled by only a portion of $\mathcal{J}^{-}$, rather than the direct effects of change in the geometry, which is already discarded in the ray-optics approximation.} This is the 
mathematical reason why the dynamics of the collapse is irrelevant to obtain the 
temperature and one obtains the same result as in the case of an eternal black hole and Hartle-Hawking vacuum state. The \textit{exact} Bogoliubov coefficients, obtained by going beyond the ray optics,   will, of course, be sensitive to the geometry change as well. Further, it is the loss of time reversal invariance during the collapse which leads to a non-zero flux of energy, which is absent in the case of an eternal black hole and Hartle-Hawking vacuum state.

A combination of these effects arise in the case of CGHS black hole in 1+1 dimension \cite{Callan:1992rs,Fabbri:2005mw}. The collapse of quantum matter leads to the standard black hole evaporation scenario with observers at $i^{+}_R (t\to\infty, x\to+\infty)$ detecting a flux of thermal radiation, which is well-known in the literature. This situation is mathematically identical to what happens in the case 1+3 spherically symmetric collapse. \emph{But it turns out that there is another effect in the same spacetime} which is unnoticed in the literature, which is conceptually very similar to Unruh effect in its structure: Observers at future left asymptotic $\mathcal{J}^{+}_L(t\to\infty, x\to-\infty)$ find themselves in a thermal environment at the same temperature as seen by observers at $\mathcal{J}^{+}_R$ (but without any associated flux!), however these observers \emph{do not have to necessarily accelerate} unlike their right moving counterparts. This result is rather curious because these are 
inertial observers in a flat region of the spacetime who see no change in geometry as they reside in the causal past of the collapsing shell. This shows one instance of Unruh effect in flat spacetime \textit{without any classical source},  which we will argue originates through \emph{quantum correlations} in analogy with the standard EPR phenomenon.

There are cases, in which a geodesic observer witnesses particle excitations in the vacuum state, e.g., thermal radiation witnessed by an inertial observer in flat spacetime in presence of a receding mirror \cite{Birrell:1982ix}. The  acceleration of the mirror modifies the boundary conditions of the quantum field, which the inertial observer will ascribe as the \textit{source} of the radiation. Other instances of geodesic observers experiencing radiation arise in special geodesic  trajectories in curved spacetime, like e.g., the $r=0$ observer in de Sitter spacetime (for another interesting example, see \cite{Emelyanov:2014owa}). But if the observer measures the curvature tensor at $r=0$ in de Sitter spacetime she will find it to be non-zero and thus will know that something nontrivial is happening to the geometry \cite{Gibbons:1977mu}. Here gravity plays the role of acceleration, as evident from the equivalence principle. 

The effect we discuss in this paper on the other hand, arises for geodesic observers in a flat region of spacetime, i.e., if the observers measure the curvature tensor in a finite neighborhood around their trajectory, they will find that it is zero. The effect arises due to oblivion of the physics in the portion of interest, to a part of the modes on past Cauchy surface (which we will be calling  $ {\cal J}^{-}_R $) and the mathematics closely parallels standard Unruh effect. But, unlike the standard Unruh effect, the necessity to trace over modes arises due to the \textit{dynamics} of collapse, which takes place in a region of the spacetime which is not causally connected to the observers! Thus, what we have is a setting where:
\begin{enumerate}

\item  A family of inertial observers in a flat portion of spacetime experience thermal ambiance in the vacuum state of the field.

\item There is no classical source, (e.g., moving mirror, a distant black hole, local curvature, acceleration, time dependent
geometry etc.) which can classically explain the source of such a thermal effect.

\end{enumerate}
This effect (which, as far as we know, has been missed in the CGHS literature) constitutes a dynamics realization of Unruh effect in flat spacetime.
\section{A Picture Book Representation}

In order to explain the concepts involved in this, rather peculiar result,  we will first provide a picture-book description of how the result arises, which should demystify it. 

\begin{figure}[h]
    \begin{center}
        \begin{tikzpicture}[scale=0.8]   
            \draw[black] (0,-3) -- (3,0);
            
            \draw[black, thin] (3,0) -- (1.5,1.5);
            \draw[black] (1.5,1.5) -- (0,3);
            \draw[black] (0,-3) -- (-3,0) -- (0,3);
                         
            \draw[magenta,thick,dashed, bend left=25] (0,-3) to (0,3);
            \draw[magenta,thick,dashed, bend right=25] (0,-3) to (0,3);
            
            \draw[green!90,line width=0.5mm,bend left=45] (-1.5,1.5) to (-1.5,-1.5);

            \draw[violet, thick] (-1.5,-1.5) -- (1.5,1.5);
            \draw[violet, thick] (-1.5,1.5) -- (1.5,-1.5);
            
            \node[label=right:$\mathcal{J}_{L'}^+$] (4) at (-4.5,0.9) {};
            \node[label=right:$\mathcal{J}_R^-$] (5) at (1.5,-1.5) {};
           
            \node[label=left:$\mathcal{J}_L^+$] (7) at (-1.5,1.5) {};
            \node[label=left:$\mathcal{J}_L^-$] (8) at (-1.5,-1.5) {};
           \node[label=right:$\mathcal{J}_R^+$] (9) at (1,2) {};
             \node[label=above: $i^{+}$] at (0,3) {};
             \node[label=below: $i^{-}$] at (0,-3) {};

           \draw[->,bend left=35] (-3.5,0.9) to (-2.3,0.8);

        \end{tikzpicture}
    \end{center}
    \caption{Rindler trajectory in Minkowski spacetime}
    \label{fig:penrose06}
\end{figure}
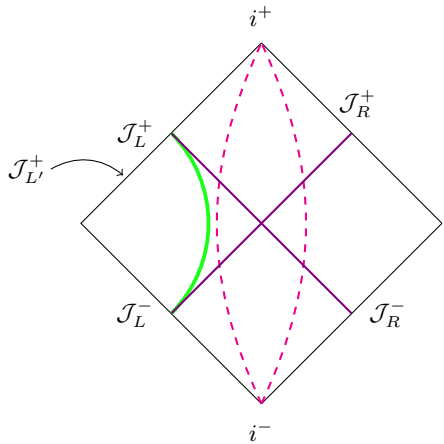

We start with the Minkowski spacetime, for which the Penrose diagram corresponds to \ref{fig:penrose06}. The full spacetime is bounded by four null lines depicting future and past left/right null infinities, viz $\mathcal{J}_{R}^{+}$, $\mathcal{J}_{R}^{-}$, $\mathcal{J}_{L}^{+}$ and $\mathcal{J}_{L}^{-}$. In this spacetime any inertial (geodesic) observers will start from past timelike infinity $i^{-}$ at $t=-\infty$ and would reach future timelike infinity $i^{+}$ at $t=\infty$. Two such geodesic observers moving leftwards and rightwards respectively, are shown by dashed curves in \ref{fig:penrose06}. In addition, there can also be some accelerated observers. An important set of such accelerated observers is the eternally accelerating Rindler observer. The trajectory of the Rindler observer starts on $\mathcal{J}_{L}^{-}$ and accelerates along a hyperbolic path to reach $\mathcal{J}^{+}_{L}$ (shown in the thick green curve in \ref{fig:penrose06}). Let us consider the past causal support of the trajectories 
of the Rindler observer vis-\'{a}-vis the inertial observer. The inertial observer has causal access to the full spacetime, whereas the spacetime region accessed by the Rindler observer is only a part of the full Minkowski spacetime. Thus the vacuum state for the inertial observer (who can access the full spacetime) would be different from that of the Rindler observer (for whom only a part of the spacetime is allowed) leading to non zero Bogoliubov coefficients, given as
\bea 
\alpha_{\Omega \omega} =\frac{1}{2\pi a} \sqrt{\frac{\Omega}{\omega}}\exp{\left[\frac{\pi\Omega}{2a}\right]}\exp{\left[-\frac{i\Omega }{a} \log{\frac{\omega}{a}} \right]}\Gamma\left[\frac{i\Omega }{a}\right],\\
\beta_{\Omega \omega} =-\frac{1}{2\pi a} \sqrt{\frac{\Omega}{\omega}}\exp{\left[-\frac{\pi\Omega}{2a}\right]}\exp{\left[-\frac{i\Omega }{a} \log{\frac{\omega}{a}} \right]}\Gamma\left[\frac{i\Omega }{a}\right], \label{BT_Rindler}
\eea
with $a$ being the acceleration of the Rindler observer  \cite{Mukhanov-Winitzky}. Therefore the Rindler observer sees the vacuum of the inertial observer, denoted by $|0\rangle _{\rm inertial}$ to be thermally occupied, i.e.,
\bea
{}_{\text{inertial}}\langle 0|N_{\Omega}|0\rangle_{\text{inertial}} = \int_{\omega} |\beta _{\Omega \omega}|^2 = \frac{1}{e^{\frac{2\pi\Omega}{a}}-1}.
\eea
where $N_{\Omega}$ stands for densitized particle number operator, essentially removing an overall volume factor from the computation. The thermal ambiance of the Rindler observer can further be verified through correlations among the various modes which can be shown to be thermal as well. In particular for a two-point correlation we obtain
\bea
{}_{\text{inertial}}\langle 0|N_{\Omega_1}N_{\Omega_2}|0\rangle_{\text{inertial}}=\left(\int_{\omega} |\beta _{\Omega_1 \omega}|^2\right)\left(\int_{\omega'} |\beta _{\Omega_2 \omega'}|^2\right) = \frac{1}{e^{\frac{2\pi\Omega_1}{a}}-1}\frac{1}{e^{\frac{2\pi\Omega_2}{a}}-1}. \label{BT_correltor}
\eea
Since we are considering a free field theory, using Wick's theorem for free fields, one can compute all the higher order correlators as well using the two point correlator and verify that these are exactly thermal \cite{Fabbri:2005mw}.

\begin{figure}[h]
    \begin{center}
        \begin{tikzpicture}[scale=0.8]   
            \draw[red!60,line width=1mm] (1.5,-1.5) -- (3,0) -- (0,3);
            \draw[red!60,line width=1mm] (3,0) -- (1.5,1.5);
            \draw[red!60,line width=1mm] (-1.5,1.5) -- (0,3);
            \draw[green,line width=1mm] (1.5,-1.5) -- (0,-3) -- (-3,0) -- (-1.5,1.5);
            \draw[black,line width=0.01mm] (1.5,-1.5) -- (3,0) -- (0,3);
            \draw[black,line width=0.01mm] (3,0) -- (1.5,1.5);
            \draw[black,line width=0.01mm] (-1.5,1.5) -- (0,3);
            \draw[black,line width=0.01mm] (1.5,-1.5) -- (0,-3) -- (-3,0) -- (-1.5,1.5);

            \draw[magenta,dashed, bend right=7] (0,-3) to (0,3);
            \draw[violet,thick,bend right=25] (0,-3) to (-1.5,1.5);
            
            \draw[black,thick,dotted,bend left=45] (-1.5,1.5) to (-1.5,-1.5);

            \draw[blue] (-1.5,-1.5) -- (1.5,1.5);
            \draw[blue] (-1.5,1.5) -- (1.5,-1.5);
            
            \node[label=right:$\mathcal{J}_{L'}^+$] (4) at (-4.5,0.9) {};
            \node[label=right:$\mathcal{J}_R^-$] (5) at (1.5,-1.5) {};
           
            \node[label=left:$\mathcal{J}_L^+$] (7) at (-1.5,1.5) {};
            \node[label=left:$\mathcal{J}_L^-$] (8) at (-1.5,-1.5) {};
           \node[label=right:$\mathcal{J}_R^+$] (9) at (1,2) {};
             \node[label=above: $i^{+}$] at (0,3) {};
           \node[label=above: III] at (-1.5,-0.5) {};
           \node[label=above: I] at (1.5,-0.5) {};
             \node[label=below: $i^{-}$] at (0,-3) {};

           \draw[->,bend left=35] (-3.5,0.9) to (-2.3,0.8);

        \end{tikzpicture}
    \end{center}
    \caption{Non-geodesic Observer in Minkowski spacetime}
    \label{fig:penrosenew01}
\end{figure}
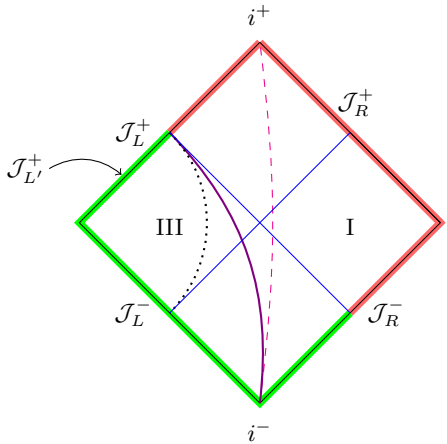

Alternatively, the observations carried out by the Rindler observer can be described using a density matrix obtained by tracing out modes inaccessible to her. If the field is in the inertial vacuum state, the resulting density matrix will be thermal.
 As an illustration let us consider \ref{fig:penrosenew01}, where three observers are shown. The dashed observer is an inertial observer and, as in the earlier figure, has access to the full spacetime;  the dotted trajectory is that of the standard, eternally accelerating  Rindler observer;
the third trajectory (represented by thick magenta line) represents an observer who originally started as inertial but then changed her mind and accelerates uniformly to end up on $\mathcal{J}_{L}^{+}$ just like the Rindler observer. This observer also has past causal connection only to the past domain of dependence of $\mathcal{J}_{L'}^{+}$, which being only a portion of the full manifold, will lead to a mixed density matrix having thermal character at late times. At finite times, its trajectory will be different from the Rindler observer and unlike the Rindler observer will have a future causal link  to the right wedge. Hence the density matrix will be different. The same can be seen through the Bogoliubov coefficients, which will be similar to those of the Rindler observer for high frequencies at late times only. The exact form of the Bogoliubov coefficients will differ for small frequencies as the large wavelengths will take them on different trajectories.

\begin{figure}[h]
    \begin{center}
        \begin{tikzpicture}[scale=0.8]   
            \draw[green,line width=1mm] (0,-3) -- (1.5,-1.5);
            \draw[black,line width=0.1mm] (0,-3) -- (1.5,-1.5);
            \draw[green,line width=1mm] (0,-3) -- (-3,0);
            \draw[black,line width=0.1mm] (0,-3) -- (-3,0);
            \draw[black,thick,dashed] (1.5,1.5) -- (-1.5,1.5) -- (0,3) -- (1.5,1.5);
            \draw[green,line width=1mm] (-3,0) -- (-1.5,1.5);
            \draw[black,line width=0.1mm] (-3,0) -- (-1.5,1.5);
            \draw[red!60,line width=1mm] (1.5,-1.5) -- (3,0) -- (1.5,1.5);
            \draw[black,line width=0.1mm] (1.5,-1.5) -- (3,0) -- (1.5,1.5);
            
            \draw[violet,line width=0.5mm,bend left=25] (0,-3) to (-1.5,1.5);

            \draw[blue] (-1.5,-1.5) -- (1.5,1.5);
            \draw[blue] (-1.5,1.5) -- (1.5,-1.5);
            \node[label=right:$\mathcal{J}_R^-$] (5) at (1.5,-1.5) {};
            \node[label=right:$\mathcal{S}$] (2) at (1.1,3.4) {};
            \node[label=right:$\mathcal{J}_{R}^+$] (4) at (3.6,1.4) {};
           
            \node[label=left:$\mathcal{J}_{L'}^+$] (7) at (-2.0,1.2) {};
            \node[label=left:$\mathcal{J}_L^-$] (8) at (-1.5,-1.5) {};

           \node[label=above: $i_R^{+}$] at (1.5,1.5) {};
             \node[label=below: $i^{-}$] at (0,-3) {};
	    \node[label=above: $i_{L}^{+}$] at (-1.5,1.5) {};
           \draw[->,bend right=35] (2) to (0,1.7);
           \draw[->,bend right=35] (4) to (2.3,0.8);

        \end{tikzpicture}
    \end{center}
    \caption{Observers in a hypothetical spacetime}
    \label{fig:penrosenew02}
\end{figure}
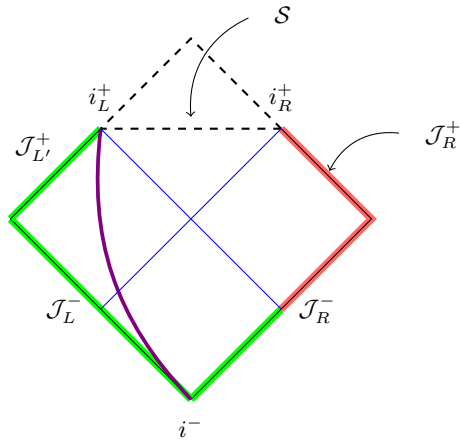

Let us now consider a hypothetical scenario in which some portion of Minkowski spacetime becomes inaccessible, as shown in  \ref{fig:penrosenew02} by the dashed triangular region. (Right now this is the Penrose diagram of some hypothetical spacetime; we will soon see how it actually arises in the CGHS case.) Further, we assume this truncation of spacetime is such that it requires  left-moving geodesic observers  to terminate on $i_{L}^{+}$ instead of $i^{+}$ as in \ref{fig:penrosenew01}. 
Since these observers derive their causal support only from the past of $\mathcal{J}_{L'}^{+}$, they are compelled to trace over a portion of field configuration on $\mathcal{J}_{R}^{-}$ and hence they will end up using a density matrix which is mixed. Whenever \emph{any} observer, \emph{geodesic or not}, who derives her past causal support from a subset of the full spacetime, the global vacuum state will appear to be non-vacuous at late times. 

The geodesic part of the above statement might appear perplexing. One may wonder how a geodesic observer can ever experience such a nontrivial effect usually associated with accelerated observers. The role of acceleration is only to make part of the spacetime region inaccessible; if we can achieve this by some other means, we will still have the same result. Indeed, there can be various other scenarios where geodetic observers can witness thermal radiation. However, in this construction, the left moving observers are unable to even associate any classical source to these radiations.

The situation as depicted in \ref{fig:penrosenew02} appears unphysical as presented --- because we have artificially removed part of the spacetime ---, but we will later discuss a situation in which a portion of spacetime is indeed dynamically denied to a \textit{geodesic observer}, very  much like in the spirit of the observer in \ref{fig:penrosenew02} (shown in thick magenta curve). 

But before we do that, we will consider another example in the next section which might make all these less surprising. This will involve the collapse of a null shell forming a black hole (see \ref{fig:penrose01}). In the case, a timelike geodesic observer at $r=0$, will remain entirely in the flat spacetime until being eaten up by the singularity and receives causal signals only from a part of $\mathcal{J}^{-}$ (see \ref{fig:penrose01} below). This study will provide us useful insights towards the constructs to be used in the later sections of the paper. 
\section{A null shell collapse}

\begin{figure}[h]
    \begin{center}
        \begin{tikzpicture}[scale=1.0]
        
            \draw[yellow,line width=1mm] (0,-3) -- (0,1.5);
            \draw[black,line width=0.01mm] (0,-3) -- (0,1.5);
            \draw[dashed,->] (0,0) -- (1.5,1.5);
            \draw[green, thick] (1.65, 1.35) -- (0,-0.3) -- (1.35,-1.65);
       
            \draw[red,thick, snake it] (0,1.5) -- (1.5,1.5);

            \draw[black] (0,-3) -- (3,0) -- (1.5,1.5);
            \draw[black] (1.5,-1.5) -- (0,0);
        
            \draw[blue] (2.25,-0.75) to (0,1.5);
            \draw[blue] (2.35,-0.65) to (0.2,1.5);
       
            \node[label=right:$i^+$] (1) at (1.2,1.8) {};
            \node[label=right:$i^-$] (2) at (0,-3) {};
            \node[label=right:$i^0$] (3) at (3.0,0) {};
             \node[label=below:$d$](4) at (1.7,-1.3) {};
            \node[label=below:$u_i$] (7) at (2.35,-0.65){};
             \node[label=below:$u_f$] (8) at (2.55,-0.45){};
             
            \node[label=right:$\mathcal{J}^+$] (5) at (3.6,1.4) {};
            \node[label=right:$\mathcal{J}^-$] (6) at (3.6,-1.4) {};

            \node[label=right:\textrm{r=0}] at (-1,0) {};
            
            \draw[->,bend right=35] (5) to (2.3,0.8);
            \draw[->,bend left=35] (6) to (2.0,-1.1);
           
        \end{tikzpicture}
    \end{center}
    \caption{Penrose Diagram For Schwarzschild null shell collapse}
    \label{fig:penrose01}
\end{figure}
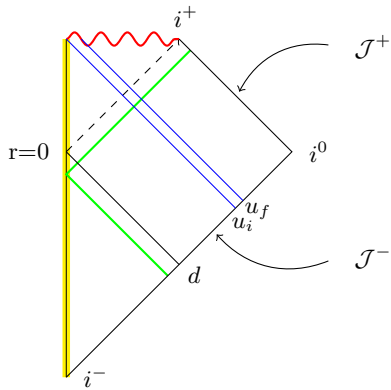

For a closer look at the above mentioned features, we will first consider a null shell collapse forming a Schwarzschild hole \cite{Birrell:1982ix,Parker:2009uva,Padmanabhan:2009vy,Singh:2014paa,Smerlak:2013sga} (see \ref{fig:penrose01}). In this case, the singularity gets originated from the co-ordinate $u=u_i$, the co-ordinate point of introduction of the null-shell. (This collapse model will be relevant for comparison with the null shell collapse in $1+1$ dimensional dilatonic gravity, forming a black hole, which is discussed later on.)
The singularity starts forming from the co-ordinate $u=u_i$, while the event horizon is located at $r=2M$, with $M$ being the mass of the shell. The co-ordinate $r=2M$ can be reflected back through $r=0$ on to $\mathcal{J}^{-}$ at $u=d$. As discussed previously, the out-going modes $v_{\omega}$ on $\mathcal{J}^{+}$ derive their past causal support completely from the portion $u<d$ on $\mathcal{J}^{-}$. Therefore the Bogoliubov coefficients of mode transformation, should be evaluated through the portion of mode functions $u_{\omega}$ in the regime $u<d$. Further, all the null rays emanating from  $\mathcal{J}^{-}$ region $u<u_{i}$ also experience a change of geometry, i.e., they start moving in flat spacetime inwards, get reflected at the origin ($r=0$) and 
then encounter the collapsing shell to feel the geometry changed into a Schwarzschild one. 

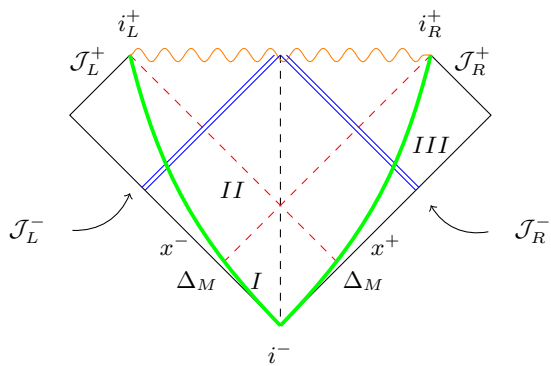
\begin{figure}[h]
    \begin{center}
        \begin{tikzpicture}[scale=0.8]
     
            \draw[dashed,->] (0,1.5) --(0,-3);
        
            \draw[orange,snake it] (-2.5,1.5) -- (2.5,1.5);
              \draw[black] (0,-3) -- (3.5,0.5) -- (2.5,1.5);
            \draw[black] (0,-3) -- (-3.5,0.5) -- (-2.5,1.5);
            \draw[blue] (2.25,-0.75) -- (0,1.5); 
            \draw[blue] (-2.25,-0.75) -- (0,1.5);
            
             \draw[blue] (2.3,-0.7) -- (0.1,1.5);
             \draw[blue] (-2.3,-0.7) -- (-0.1,1.5);
              
            \draw[red, dashed]  (2.5,1.5) -- (-1,-2);
            \draw[red, dashed]  (-2.5,1.5) -- (1,-2);
              
            \draw[green,line width=0.5mm,bend left=15] (0,-3) to (-2.5,1.5);
            \draw[green, line width=0.5mm,bend right=15] (0,-3) to (2.5,1.5);


            \node[label=right:$\mathcal{J}_R^+$] (4) at (2.6,1.4) {};
            \node[label=right:$\mathcal{J}_R^-$] (5) at (3.6,-1.4) {};
            \node[label=below: $x^-$](6) at (-1.75,-1.2) {};
            \node[label=left:$\mathcal{J}_L^+$] (7) at (-2.6,1.4) {};
            \node[label=left:$\mathcal{J}_L^-$] (8) at (-3.6,-1.4) {};
             \node[label=below: $x^+$](10) at (1.75,-1.2) {};
            \node[label=left : $\Delta_M$] at (-0.75,-2.25) {};
            \node[label=right : $\Delta_M$] at (0.75,-2.25) {};
            \node[label=center : $I$] at (-0.40,-2.25) {};
            \node[label=center : $II$] at (-0.80,-0.75) {};
            \node[label=center : $III$] at (2.50,0.0) {};
            \node[label=above: $i_L^{+}$] at (-2.5,1.5) {};
            \node[label=above: $i_R^{+}$] at (2.5,1.5) {};
             \node[label=below: $i^{-}$] at (0,-3) {};
            \draw[->,bend left=35] (5) to (2.5,-1.0);
            \draw[->,bend right=35] (8) to (-2.5,-0.8);

        \end{tikzpicture}
    \end{center}
    \caption{Schwarzschild in isotropic co-ordinates}
    \label{fig:penrose02}
\end{figure}

This can be more clearly seen in an isotropic co-ordinate system (see \ref{fig:penrose02}) which uses a Cartesian $x$ coordinate with $-\infty<x<\infty$ instead of the usual radial coordinate $r$ with $0<r<\infty$. The relevant Penrose diagram is shown in \ref{fig:penrose02}.
In these coordinates, (see \ref{fig:penrose02}) there will be two past null infinities, namely left $(\mathcal{J}_L^-)$ and right $(\mathcal{J}_R^-)$, as well as two future null infinities $\mathcal{J}_L^+$ and $\mathcal{J}_R^+$. The location of event horizon in these co-ordinates is marked to (say) $\Delta_M$, which corresponds to $r=2M$ in Schwarzschild co-ordinates. It is usual to  define the modes as left moving or right moving in these co-ordinates. A right moving mode originates from $\mathcal{J}_L^-$ and ends up on $\mathcal{J}_R^+$ and vice versa for the left-moving modes. Due to spherical symmetry, consideration of any set of past and future asymptotic observer pairs will be equivalent to any other. A time-like observer (thick green curves in \ref{fig:penrose02}), similarly starting from past time like infinity and escaping the black hole region, either end up on $\textit{i}^{~+}_L$ or on $\textit{i}^{~+}_R$ depending upon whether the observer moves leftwards or rightwards. 

Let us consider a set of modes moving rightwards. Any null ray originating from $\mathcal{J}_L^-$ and ending up on $\mathcal{J}_R^+$ will  experience a change of geometry after it crosses  the collapsing null shell. For observers on $\mathcal{J}_R^+$, the isotropic co-ordinate $|x^{-}| =\Delta_M $ marks the location of event horizon. Therefore, the modes reaching $\mathcal{J}_R^+$ derive their past causal support from the region $I$ in \ref{fig:penrose02}. No event in the region $II$ is connected to $\mathcal{J}_R^+$. An exactly similar picture is there for the left-moving modes. 

In the standard black hole analysis, the out-going mode functions on $\mathcal{J}_R^+$ are given as 
\bea
v_{\omega} = \frac{1}{\sqrt{2 \omega}} e^{-i \omega u}; \hspace{0.5 in} u \in (-\infty, \infty). \label{FutureModes01}
\eea
These mode functions provide an orthonormal basis across complete $\mathcal{J}_R^+$. Similarly the right-moving modes on $\mathcal{J}_L^-$ are spanned by
\bea
u_{\omega} = \frac{1}{\sqrt{2 \omega}} e^{-i \omega v}; \hspace{0.5 in} v \in (-\infty, \infty). \label{PastModes01}
\eea
The Bogoliubov transformation coefficients between these modes can be evaluated by taking the covariant inner products on $\mathcal{J}_L^-$. However, in order to do that we need to express mode functions $v_{\omega}$ on $\mathcal{J}_L^-$  in terms of $u_{\omega}$. For that purpose we need to track the out-going modes at $\mathcal{J}_R^+$ all the way down to $\mathcal{J}_L^-$. In principle, this a formidable job, since the exact form of modes in the whole spacetime is complicated at best, if obtainable in closed form. However, appealing to ray-optics approximation \cite{Hawking:1974sw,Parker:2009uva} for modes very close to the horizon, we in a sense, avoid this issue. This approximation gives us the expression of modes $v_{\omega}$ close to $|x^{-}| =\Delta_M $, i.e., $u=d$  in terms of $u_{\omega}$. We obtain the Bogoliubov coefficients readily as
\bea
\alpha_{\Omega \omega} = -2 i \int_{-\infty}^{0} d x^{-} u_{\Omega} \partial_{-}u^{*}_{\omega}, \nonumber\\
\beta_{\Omega \omega} = 2 i \int_{-\infty}^{0} d x^{-} u_{\Omega} \partial_{-}u_{\omega}.
\eea
However, as we discussed this integration has to be truncated to within the region $u<d$, which gives it a Rindler kind of appearance, making it \cite{Hawking:1974sw,Birrell:1982ix,Parker:2009uva}
\bea 
\alpha_{\Omega \omega} &=&\frac{1}{2\pi \kappa} \sqrt{\frac{\Omega}{\omega}}\exp{\left[\frac{\pi\Omega}{2\kappa}\right]}\exp{[i(\Omega-\omega)d]}\exp{\left[\frac{i\Omega }{\kappa} \log{\frac{\omega}{C}} \right]}\Gamma\left[-\frac{i\Omega }{\kappa}\right],\nonumber\\
\beta_{\Omega \omega} &=&-\frac{1}{2\pi \kappa} \sqrt{\frac{\Omega}{\omega}}\exp{\left[-\frac{\pi\Omega}{2\kappa}\right]}\exp{[i(\Omega+\omega)d]}\exp{\left[\frac{i\Omega }{\kappa} \log{\frac{\omega}{C}} \right]}\Gamma\left[-\frac{i\Omega }{\kappa}\right]. \label{BT}
\eea
where $C=C_1C_2$ is a product of affine parameters for in-going ($C_1$) and out-going ($C_2$) null geodesics with $\kappa =1/4M$ being the surface gravity associated with the black hole horizon. Subsequent computation of the expectation value of densitized number operator in the `in' vacuum yields, 
\begin{align}
_{\rm in}\langle 0|N_{\Omega}|0\rangle _{\rm in}=\int _{\omega}|\beta _{\Omega \omega}|^{2}=\frac{1}{\exp(\frac{2\pi \Omega}{\kappa})-1}
\end{align}
which is definitely thermal. Along identical lines one can immediately verify that the higher order correlations will also be thermal in nature and justifying the thermal ambiance of black holes. From the rather general nature of the analysis we expect this result to give the Bogoliubov coefficients for the timelike geodesic observer stationary at $r=0$ if we replace $d$ by $u_{i}$. Such an observer is engulfed by a \emph{true}  singularity when the shell collapse to $r=0$, it results in geodesic incompleteness, something we will come back to, in a subsequent work.

In the limit $M\rightarrow 0$, the portion being denied causally, becomes small. Further  the geometry change as well as the formation of the singularity does not occur, making $\mathcal{J^{+}}$ a Cauchy surface, which is not the case when $M\neq0$. Since the effect of causal denial is intimately tied with the effect of geometry change, it is difficult to account for these effects individually in a general case in 1+3 dimension. Nevertheless, one can argue that the ``tracing over'' of modes, which are causally not accessible, alone leads to Unruh effect with zero flux, whereas the geometry change makes the thermal radiation more `real' with a non-zero flux \cite{Davies:1976ei,Ford:1993bw,Ford:1994bj,Ford:1995gb}. The non vanishing of flux can  be associated with moving of the geometry away from being a flat one. We will see below that these two effects can nicely be segregated in a $1+1$ dimensional collapse model in dilaton gravity.
\section{(1+1)-dimensional dilatonic black hole}\label{dilaton}

The CGHS black hole \cite{Callan:1992rs,Fabbri:2005mw} is a $1+1$ dimensional gravitational collapse model of a dilatonic field  $\phi$ interacting with gravity in the presence of cosmological constant $\lambda$ and matter fields $f_i$, described by the action,
\begin{align}
\mathcal{A}=
\frac{1}{2\pi}\int d^{2}x\sqrt{-g}\left[e^{-2\phi}\left(R+4(\mathbf{\nabla} \phi)^{2}+4\lambda ^{2}\right)-\frac{1}{2}\sum _{i=1}^{N}(\mathbf{\nabla}f_{i})^{2} \right]. \label{CGHS_Action}
\end{align}
This action, when compared with the standard Hilbert Einstein action, can be viewed as written in a conformally related frame. In standard Einstein frame, obtained through conformal transformation, the dilaton field makes appearance as a canonical scalar field.
Since all two dimensional space-times are conformally flat the metric ansatz will involve a single unknown function, the conformal factor $\rho$, which is written in double null coordinates as,
\begin{align}
ds^{2}=-e^{2\rho}dx^{+}dx^{-}, \text{ with,}\\
x^{+} \in (0,\infty), \text{ and } x^{-} \in (-\infty, 0). \nonumber
\end{align}
For the matter fields, the classical solutions are those in which, $f_{i}=f_{i+}(x^{+})+f_{i-}(x^{-})$. Given some particular matter
fields one can obtain corresponding solutions for $\phi$ and $\rho$ respectively from the equations of motion. A simple static solution corresponds to  $e^{-2\rho}=e^{-2\phi}=(M/\lambda)-\lambda ^{2}x^{+}x^{-}$, representing a black hole of mass $M$, with a line element 
\begin{align}
ds^{2}=-\frac{dx^{+}dx^{-}}{\frac{M}{\lambda}-\lambda^2 x^{+}x^{-}}. \label{LineElement}
\end{align}
In absence of the mass, $M=0$ and we obtain {\it linear dilaton vacuum} line element from \ref{LineElement}.

We consider, for the collapsing scenario, a simplistic case where only one scalar field $f$ is present. The matter moving leftwards collapses to form a black hole \cite{Callan:1992rs,Mikovic:1996bh}. If the matter distribution starts at $x_{i}^{+}$ and extends up to $x_{f}^{+}$, then the line element corresponding to this matter configuration turns out to be,
\begin{align}
ds^{2}=-\frac{dx^{+}dx^{-}}{\frac{M(x^{+})}{\lambda}-\lambda ^{2}x^{+}x^{-}-P^{+}(x^{+})x^{+}}, \label{CGHSMetric}
\end{align}
where the functions $M(x^{+})$ and $P^{+}(x^{+})$ correspond to the integrals,
\bea
M(x^{+})&=&\int _{x_{i}^{+}}^{x^{+}}dy^{+}y^{+}T_{++}(y^{+}), \label{MassInt}\\
P^{+}(x^{+})&=&\int _{x_{i}^{+}}^{x^{+}}dy^{+}T_{++}(y^{+}).
\eea
The  matter field  satisfies the null energy condition and hence the quantity defining the mass $M(x^{+})$ remains positive semi-definite.
The region outside $x_{f}^{+}$ is a black hole of mass $M\equiv M(x_{f}^{+})$ \cite{Fabbri:2005mw}. There is a curvature singularity at $e^{-\rho} = 0$.
\begin{figure}[h]
    \begin{center}
        \begin{tikzpicture}[scale=0.8]
     
            \draw[dashed,->] (-1.5,-1.5) -- (0,0) -- (1.5,1.5);
        
            \draw[orange, snake it] (-1.5,1.5) -- (1.5,1.5);
              \draw[black] (0,-3) -- (3,0) -- (1.5,1.5);
            \draw[black] (0,-3) -- (-3,0) -- (-1.5,1.5);
            \draw[blue] (1.5,-1.5) -- (0,0) -- (-1.5,1.5); 
             \draw[blue] (1.60,-1.40) -- (-1.3,1.50);
             
            \draw[magenta,line width=0.5mm,bend right=15] (0,-3) to (1.5,1.5);


            \node[label=right:$\mathcal{J}_R^+$] (4) at (3.6,1.4) {};
            \node[label=right:$\mathcal{J}_R^-$] (5) at (3.6,-1.4) {};
            \node[label=below: $x_i^+$](6) at (1.5,-1.4) {};
            \node[label=below: $x^-$](6) at (-0.75,-2.2) {};
            \node[label=left:$\mathcal{J}_L^+$] (7) at (-3.6,1.7) {};
            \node[label=left:$\mathcal{J}_L^-$] (8) at (-3.6,-1.4) {};
             \node[label=below: $x_f^+$](9) at (2.1,-0.9) {};
             \node[label=below: $x^+$](10) at (0.75,-2.2) {};
              \node[label=above: $i_L^{+}$] at (-1.5,1.5) {};
            \node[label=above: $i_R^{+}$] at (1.5,1.5) {};
             \node[label=below: $i^{-}$] at (0,-3) {};
             \node[label=below: $\mathcal{H}_{R}$] at (-0.5,-0.5) {};           

            \draw[->,bend right=35] (4) to (2.3,0.8);
            \draw[->,bend left=35] (5) to (2.5,-1.0);
            \draw[->,bend left=35] (7) to (-2.3,0.8);
            \draw[->,bend right=35] (8) to (-2.5,-0.8);

        \end{tikzpicture}
    \end{center}
    \caption{Penrose Diagram for a CGHS black hole}
    \label{fig:penrose03}
\end{figure}
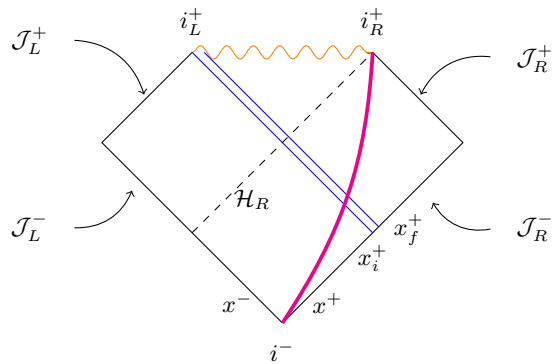
The singularity hides behind an event horizon (which is located at $x^{-}=-P^{+}/\lambda ^{2}$) for future null observers receiving the out-moving radiation. The location of the event horizon can be obtained starting from the location of the apparent horizon, see \ref{fig:penrose03}. This can be obtained using $\partial _{+}A\leq 0$, where $A$ stands for the transverse area of the horizon in a $3+1$ dimensional setting. Borrowing this idea to $1+1$ dimension, this equality would lead to the location of the event horizon at $x^{-}=-P^{+}/\lambda ^{2}$, in the out region, i.e., after $x^{+}>x^{+}_{f}$.

Thermodynamics as well as Hawking evaporation of such black hole solutions have been extensively studied \cite{Fabbri:2005mw, Birnir:1992by, Giddings:1992ff, Alves:1995hn, Alves:1999ct, Mikovic:1996bh}. We introduce a co-ordinate set $z^{\pm}$ suited for the in-region $\mathcal{J}_{L}^{-}$
\bea
\pm \lambda x^{\pm} = e^{\pm \lambda z^{\pm}},
\eea
which maps the entire $\mathcal{J}_{L}^{-}$ into $z^{-}\in(-\infty,\infty)$. We also introduce another co-ordinate system suited for $\mathcal{J}_{R}^{+}$ as $\sigma _{\rm out}^{\pm} \in(-\infty,\infty)$ where the transformation between $(z^{+},z^{-})$ and $(\sigma _{\rm out}^{+},\sigma _{\rm out}^{-})$  is given by:
\begin{align}
z^{+}=\sigma _{\rm out}^{+};\qquad z^{-}=-\frac{1}{\lambda}\ln \left[e^{-\lambda \sigma _{\rm out}^{-}}+\frac{P^{+}}{\lambda} \right].
\end{align}
The horizon located at $x^{-}=-P^{+}/\lambda ^{2}$, will get mapped to $z^{-}=z_i^{-}= -\frac{1}{\lambda}\log{(P^{+}/\lambda)}$ in these co-ordinates. `In-' state modes are defined on the asymptotically flat region $\mathcal{J}_{L}^{-}$ moving towards $\mathcal{J}_{R}^{+}$ and the convenient basis modes defined  correspond to,
\begin{align}
u_{\omega}=\frac{1}{\sqrt{2\omega}}e^{-i\omega z^{-}},
\end{align}
where $\omega >0$. The `out' region corresponds to $\mathcal{J}_{R}^{+}$ which receives the state from $\mathcal{J}_{L}^{-}$ after the black hole has formed. The basis modes in the out region at $\mathcal{J}_{R}^{+}$ are
\begin{align}
v_{\omega}(\sigma _{\rm out}^{-})&=\frac{1}{\sqrt{2\omega}}e^{-i\omega \sigma _{\rm out}^{-}}; \\
v_{\omega}(z^{-})&=\frac{1}{\sqrt{2\omega}}e^{-i\omega \sigma _{\rm out}^{-}(z^{-})}\Theta (z_i^{-}-z^{-}),
\end{align}
where $\Theta$ is the usual step function marking the fact that the out modes are supported by states on $\mathcal{J}_{L}^{-}$ between the region $(-\infty,z_i^{-})$ only. These mode functions provide a complete orthonormal basis on $\mathcal{J}_{R}^{+}$ in terms of $\sigma _{\rm out}^{-}$, however. Again the field can be specified fully on $\mathcal{J}_{L}^{-}$ or jointly on $\mathcal{J}_{R}^{+}$ and the event horizon ${\cal H}_R$. Since the mode functions at ${\cal H}_R$ correspond to part of the field falling into the singularity and such interior modes cannot be observed by observers at $\mathcal{J}_{R}^{+}$, they need to be traced over. Thus, the precise form of mode decomposition on ${\cal H}_R$ does not affect physical results for $\mathcal{J}_{R}^{+}$. Therefore, we can expand the dilaton field in different mode basis as,
\begin{align}
f&=\int _{0}^{\infty}d\omega ~\left[a_{\omega}u_{\omega}+a_{\omega}^{\dagger}u_{\omega}^{*}\right],~~~(\textrm{in})
\\
&=\int _{0}^{\infty}d\omega ~\left[b_{\omega}v_{\omega}+b_{\omega}^{\dagger}v_{\omega}^{*}+\hat{b}_{\omega}\hat{v}_{\omega}+\hat{b}_{\omega}^{\dagger}\hat{v}_{\omega}^{*}\right],~~~(\textrm{out})
\end{align}
where $a_{\omega}^{\dagger}$ corresponds to creation operator appropriate for the `in' region. Similarly $b_{\omega}^{\dagger}$ and $\hat{b}_{\omega}^{\dagger}$ stand for the creation operators for the `out' region and the black hole interior region respectively. The inner product between $v_{\Omega}$ and $u_{\omega}^{*}$ corresponds to,
\begin{align}
\alpha _{\Omega \omega}&=-\frac{i}{\pi}\int _{-\infty}^{z_i^{-}}dz^{-}v_{\Omega}\partial _{-}u_{\omega}^{*}
=\frac{1}{2\pi}\sqrt{\frac{\omega}{\Omega}}\int _{-\infty}^{z_i^{-}}dz^{-}\exp \left[\frac{i\Omega}{\lambda}\ln \left\lbrace \left(e^{-\lambda z^{-}}-\frac{P^{+}}{\lambda}\right) \right\rbrace+i\omega z^{-} \right]
\nonumber
\\
&=\frac{1}{2\pi \lambda}\sqrt{\frac{\omega}{\Omega}}\left(\frac{P^{+}}{\lambda}\right)^{i(\Omega-\omega)/\lambda}~B\left(-\frac{i\Omega}{\lambda}+\frac{i\omega}{\lambda},1+\frac{i\Omega}{\lambda} \right),\label{AlphaCGHS}
\end{align}
while the inner product between  $v_{\Omega}$ and $u_{\omega}$ gives
\begin{align}
\beta _{\Omega \omega}&=\frac{i}{\pi}\int _{-\infty}^{z_i^{-}}dz^{-}v_{\Omega}\partial _{-}u_{\omega}
=\frac{1}{2\pi}\sqrt{\frac{\omega}{\Omega}}\int _{-\infty}^{z_i^{-}}dz^{-}\exp \left[\frac{i\Omega}{\lambda}\ln \left\lbrace \left(e^{-\lambda z^{-}}-\frac{P^{+}}{\lambda}\right) \right\rbrace-i\omega z^{-} \right]
\nonumber
\\
&=\frac{1}{2\pi \lambda}\sqrt{\frac{\omega}{\Omega}}\left(\frac{P^{+}}{\lambda}\right)^{i(\Omega+\omega)/\lambda}~B\left(-\frac{i\Omega}{\lambda}-\frac{i\omega}{\lambda},1+\frac{i\Omega}{\lambda} \right), \label{BetaCGHS}
\end{align}
with $B(x,y)$ being the Beta function. Therefore, we can verify that the late time right moving observers (such a observer is depicted by the thick magenta curve in \ref{fig:penrose03}) do obtain a thermal spectrum \cite{Fabbri:2005mw} with a temperature $\lambda/2\pi$, such that the expectation value of the densitized number operator in the `in' vacuum becomes,
\bea
_{\rm in}\langle 0|N_{\Omega}|0\rangle _{\rm in}=\int_{\omega} |\beta _{\Omega \omega}|^2
=\frac{1}{\exp(\frac{2\pi \Omega}{ \lambda})-1}, \label{NumberOperator}
\eea
which is certainly thermal. Similarly the higher correlations are also thermal with identical temperature. In two dimensions, the parameter $\lambda$ has dimension $L^{-1}$ and is a legitimate quantity to set the scale of the frequencies of emission. Further, $\lambda$ also turns out to be the surface gravity of the black hole. Interestingly, the effect of mass of the black hole can be eliminated in the redefinition of the co-ordinates in the region exterior to the black hole \cite{Fabbri:2005mw}. As a result, $M$ does not explicitly appear in the expression for the temperature, which is essentially related to the surface gravity.

Further, all references to the matter content which formed the black hole, appears only in the phase  (see \ref{AlphaCGHS} and \ref{BetaCGHS}), through $P^{+}$ and gets wiped out when we take  the modulus. If we follow a null ray originating from $\mathcal{J}_{L}^{-}$ or a timelike trajectory and moving rightwards, it suffers a change of geometry once it crosses the collapsing null shell, as in the case for Schwarzschild hole formation. Therefore, the experiences of these observers are inclusive of both tracing over and the geometry change. Hence, as associated with a black hole formation, there is an associated flux of radiation as measured by the observers attached to such trajectories \cite{Fabbri:2005mw,Mikovic:1996bh}. However, this model being asymmetric under left-right exchange, provides an opportunity of studying the effect of tracing over separately, if we consider the left-moving modes, 
which we do next.
\section{Dynamically generated Unruh effect for inertial observers in CGHS Black Hole Spacetime}

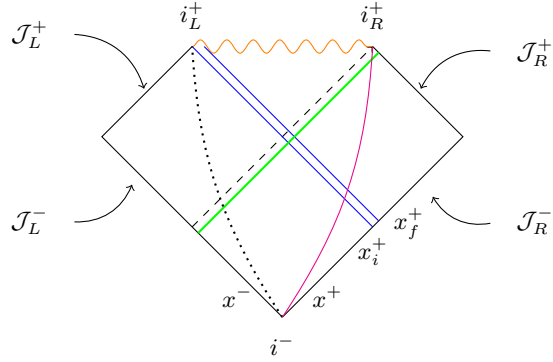
\begin{figure}[h]
    \begin{center}
        \begin{tikzpicture}[scale=0.8]
     
            \draw[dashed,->] (-1.5,-1.5) -- (0,0) -- (1.5,1.5);
        
            \draw[orange, snake it] (-1.5,1.5) -- (1.5,1.5);
              \draw[black] (0,-3) -- (3,0) -- (1.5,1.5);
            \draw[black] (0,-3) -- (-3,0) -- (-1.5,1.5);
            \draw[blue] (1.5,-1.5) -- (0,0) -- (-1.5,1.5); 
             \draw[blue] (1.60,-1.40) -- (-1.3,1.50);
             
             \draw[green, thick] (-1.4,-1.6) -- (1.6,1.4);
             
            \draw[black, dotted, thick, bend left=15] (0,-3) to (-1.5,1.5);
            \draw[magenta, bend right=15] (0,-3) to (1.5,1.5);


            \node[label=right:$\mathcal{J}_R^+$] (4) at (3.6,1.4) {};
            \node[label=right:$\mathcal{J}_R^-$] (5) at (3.6,-1.4) {};
            \node[label=below: $x_i^+$](6) at (1.5,-1.4) {};
            \node[label=below: $x^-$](6) at (-0.75,-2.2) {};
            \node[label=left:$\mathcal{J}_L^+$] (7) at (-3.6,1.7) {};
            \node[label=left:$\mathcal{J}_L^-$] (8) at (-3.6,-1.4) {};
             \node[label=below: $x_f^+$](9) at (2.1,-0.9) {};
             \node[label=below: $x^+$](10) at (0.75,-2.2) {};
           \node[label=above: $i_L^{+}$] at (-1.5,1.5) {};
            \node[label=above: $i_R^{+}$] at (1.5,1.5) {};
             \node[label=below: $i^{-}$] at (0,-3) {};

            \draw[->,bend right=35] (4) to (2.3,0.8);
            \draw[->,bend left=35] (5) to (2.5,-1.0);
            \draw[->,bend left=35] (7) to (-2.3,0.8);
            \draw[->,bend right=35] (8) to (-2.5,-0.8);

        \end{tikzpicture}
    \end{center}
    \caption{Left moving modes in a CGHS black hole}
    \label{fig:penrose04}
\end{figure}

The CGHS model is not symmetric under left-right exchange, and hence the experiences of null ray originating from $\mathcal{J}_{R}^{-}$ or a timelike trajectory moving leftwards, will be  different from what we discussed above. Such trajectories do not suffer any change in geometry in their course, hence, as we will see, the only effect a late time observer (the dotted curve in \ref{fig:penrose04}) on  $\mathcal{J}_{L}^{+}$ finds, is rooted only in the tracing over a part of Cauchy surface $\mathcal{J}_{R}^{-}$. Therefore, these observers will also obtain a thermal expectation value in \ref{NumberOperator} as we will see, but there will be no associated flux for these thermal spectrum.
The spacetime metric in the region $x^+ < x_i^+$ is given as flat, given by \ref{CGHSMetric} with $M,P^{+} \rightarrow 0$, which on using the co-ordinates
\begin{align}
x^{+}=-\frac{1}{\lambda y^{+}};\qquad x^{-}=-\frac{1}{\lambda y^{-}},\label{coordinateSet01}
\end{align}
can be written as
\bea
ds^2 =- \frac{d y^+ d y^-}{-\lambda^2 y^{+}y^{-}}.
\eea
The singularity curve originates from the co-ordinate  of the point of matter introduction, i.e., from $x_i^+$ which is marked through \ref{coordinateSet01} as $y_i^+ =-1/\lambda x_i^+$. Under another set of co-ordinate transformations, the metric on $\mathcal{J}_{R}^{-}$ can also be brought into flat form. On  $\mathcal{J}_{R}^{-}$ we adopt
\bea
e^{-\lambda \chi^+} &=& - y^+, \nonumber\\
e^{\lambda \chi^-}  &=&  y^-,
\eea
such that the metric becomes 
\bea
ds^2 = - d \chi^+ d\chi^-, \label{FlatMetric}
\eea
with $\chi^{\pm} \in (-\infty,\infty)$. Given the spacetime metric as in \ref{FlatMetric} we can construct a co-ordinate system $\{x \equiv(T,X)\}$
\bea
\chi^+ = T +  X \nonumber\\
\chi^- = T - X
\eea
to write \ref{FlatMetric} as
\bea
ds^2 = - dT^2 + dX^2
\eea
where, $T$ plays the role of inertial time coordinate in this spacetime. The left moving inertial observer is given by the trajectory 
$$\frac{d T}{d \tau}=\gamma; \hspace{0.2 in} \frac{d X}{d \tau}=-v\gamma,$$ 
where $v$ is the inertial velocity and $\gamma$ is the corresponding Lorentz factor. It is clear that in flat spacetime such an observer will have zero two-acceleration. We note that an inertial observer sitting at $X =-a$, with $a\gg 1$ will have a trajectory closely hugging ${\cal J}_L^-$ and ${\cal J}_L^+$.

Therefore the complete set of  left-moving mode functions corresponding to the field $ f_+(\chi^+)$ can be written in these co-ordinates as
\bea
u^+_{\omega}(\chi^+) = \frac{1}{\sqrt{2 \omega}}e^{-i \omega \chi^+}.
\eea
Whereas on $\mathcal{J}_{L}^{+}$ we adopt to
\bea
e^{-\lambda \tilde{\chi}^+} &=&- (y^+-y_i^+),\nonumber\\
e^{\lambda \tilde{\chi}^-} &=&  y^-, \label{coordinateSet02}
\eea
such that these new co-ordinates range in $\tilde{\chi}^{\pm} \in (-\infty,\infty)$ in the region $x^+ < x_i^+$. The metric in these new co-ordinates becomes
\bea
ds^2 =-\frac{d\tilde{\chi}^{+}d\tilde{\chi}^{-}}{1-y_i^+ e^{\lambda \tilde{\chi}^+}}. \label{ConformallyFlat}
\eea
Given the metric in the $(\tilde{\chi}^{+},\tilde{\chi}^{-})$ coordinates, one can simply solve the field equation for the minimally coupled massless scalar field in these co-ordinates using conformal invariance of the matter field in 2 dimensions. The resulting complete set of left-moving modes corresponding to the field $ f_+(\tilde{\chi}^+)$ are again given as 
\bea  
v^+_{\omega}(\tilde{\chi}^+) = \frac{1}{\sqrt{2 \omega}}e^{-i \omega \tilde{\chi}^+}.
\eea
Notably, the timelike observers corresponding to these null co-ordinates $ \tilde{\chi}^{\pm} = \tilde{T}\pm\tilde{X}$ will find the metric to be dynamic, but will be well suited observers to define positive frequency modes of the field. Thus, the relevant questions to ask to the inertial observers are how many particle modes (i.e. $e^{-i \omega T}$ excitations) were there at remote past and how many positive frequency {\it particle} modes (i.e. $e^{-i \omega \tilde{T}})$ excitations are present at late times? 

Clearly, on ${\mathcal{J}}_{R}^{-}, v^+_{\omega}$ has support only in the region $x^+<x_i^+$. The point $x_i^+$ is mapped to $\chi^+ \equiv \chi_i^+= -\frac{1}{\lambda} \log{(-y_i^+)}$. Therefore, the Bogoliubov transformation coefficients between these two set of observers can be obtained exactly as in \ref{AlphaCGHS}, \ref{BetaCGHS} but with the replacement $|y_i^+| \leftrightarrow P^{+}/\lambda$ as
\begin{align}
\alpha _{\Omega \omega}&=-\frac{i}{\pi}\int _{-\infty}^{\chi_i^+}d\chi^+v_{\Omega}\partial _{-}u_{\omega}^{*}
=\frac{1}{2\pi}\sqrt{\frac{\omega}{\Omega}}\int _{-\infty}^{\chi_i^+}d\chi^+\exp \left[\frac{i\Omega}{\lambda}\ln \left\lbrace\left(e^{-\lambda \chi^+}-|y_i^+| \right) \right\rbrace+i\omega \chi^+ \right]
\nonumber
\\
&=\frac{1}{2\pi \lambda}\sqrt{\frac{\omega}{\Omega}}|y_i^+|^{\frac{i(\Omega-\omega)}{\lambda}}~B\left(-\frac{i\Omega}{\lambda}+\frac{i\omega}{\lambda},1+\frac{i\Omega}{\lambda} \right),\label{AlphaCGHS-R-L}
\end{align}
while 
\begin{align}
\beta _{\Omega \omega}&=\frac{i}{\pi}\int _{-\infty}^{\chi_i^+}d\chi^+v_{\Omega}\partial _{+}u_{\omega}
=\frac{1}{2\pi}\sqrt{\frac{\omega}{\Omega}}\int _{-\infty}^{\chi_i^+}d\chi^+\exp \left[\frac{i\Omega}{\lambda}\ln \left\lbrace  \left(e^{-\lambda \chi^+}-|y_i^+|\right) \right\rbrace-i\omega \chi^+ \right]
\nonumber
\\
&=\frac{1}{2\pi \lambda}\sqrt{\frac{\omega}{\Omega}}|y_i^+|^{\frac{i(\Omega+\omega)}{\lambda}}~B\left(-\frac{i\Omega}{\lambda}-\frac{i\omega}{\lambda},1+\frac{i\Omega}{\lambda} \right). \label{BetaCGHS-R-L}
\end{align}
This form of the Bogoliubov coefficients results from the integration of the positive frequency modes over $\mathcal{J}_{\rm L}^{+}$, having support in the past only on a portion of $\mathcal{J}_R^{-}$. The horizon for $\mathcal{J}_L^{+}$ is given as $y^{+} =y_{i}^{+}$, which also marks the point of singularity. More importantly, if one is interested in the late time response, i.e., the observers reaching $i_{\rm L}^{+}$, we need to take the high frequency behaviour $(\omega/\lambda \gg 1)$ of the Bogoliubov coefficients
\begin{align}
\alpha _{\Omega \omega}&\longrightarrow \frac{1}{2\pi \lambda}\sqrt{\frac{\omega}{\Omega}}|y_i^+|^{\frac{i(\Omega-\omega)}{\lambda}}\exp{\left[\frac{\pi\Omega}{2\lambda}\right]}\exp{\left[-\frac{i\Omega }{\lambda} \log{\frac{\omega}{\lambda}} \right]}\Gamma\left[\frac{i\Omega }{\lambda}\right]; \nonumber\\
\beta _{\Omega \omega} &\longrightarrow -\frac{1}{2\pi \lambda}\sqrt{\frac{\omega}{\Omega}}|y_i^+|^{\frac{i(\Omega+\omega)}{\lambda}}\exp{\left[-\frac{\pi\Omega}{2\lambda}\right]}\exp{\left[-\frac{i\Omega }{\lambda} \log{\frac{\omega}{\lambda}} \right]}\Gamma\left[\frac{i\Omega }{\lambda}\right],
\end{align}
which are (upto overall phases) exactly the same as for the Rindler observer \ref{BT_Rindler}. The overall phase factors are not really important for identifying the field content, as seen in \ref{BT_correltor}. Therefore, just like the Rindler observer, the left-moving observer at high frequencies, at late times sees the `in' vacuum  to be thermally occupied
\bea
{}_{\text{in}}\langle 0|N_{\Omega}|0\rangle_{\text{in}} = \int_{\omega} |\beta _{\Omega \omega}|^2 = \frac{1}{e^{\frac{2\pi\Omega}{\lambda}}-1}.
\eea
Again, as before, the thermal ambiance of such an observer can be verified through correlations among the various modes which also turn out to be thermal in these settings,
\bea
{}_{\text{in}}\langle 0|N_{\Omega_1}N_{\Omega_2}|0\rangle_{\text{in}}=\left(\int_{\omega} |\beta _{\Omega_1 \omega}|^2\right)\left(\int_{\omega'} |\beta _{\Omega_2 \omega'}|^2\right) = \frac{1}{e^{\frac{2\pi\Omega_1}{\lambda}}-1}\frac{1}{e^{\frac{2\pi\Omega_2}{\lambda}}-1}.
\eea
Similarly, any observable can be evaluated in the `in' vacuum using these Bogoliubov coefficients and the result will turn out to be as for a thermal state. Thus, the experiences of the geodesic observer start resembling that of a Rindler observer and the observer finds the field content to be thermal (albeit at a temperature $\lambda/2\pi$).

\begin{figure}[h]
    \begin{center}
        \begin{tikzpicture}[scale=0.8]
     
            \draw[red!60,line width=1mm] (-1.2,-1.8) -- (1.8,1.2);
	    \draw[black,line width=0.01mm] (-1.2,-1.8) -- (1.8,1.2);
            \draw[orange, snake it] (-1.5,1.5) -- (1.5,1.5);
              \draw[black] (0,-3) -- (3,0) -- (1.5,1.5);
            \draw[black] (0,-3) -- (-3,0) -- (-1.5,1.5);
            \draw[blue] (1.5,-1.5) -- (0,0) -- (-1.5,1.5); 
             \draw[blue] (1.60,-1.40) -- (-1.3,1.50);
             
            \draw[magenta, bend left=15] (0,-3) to (-1.5,1.5);
            \draw[magenta, bend right=15] (0,-3) to (1.5,1.5);

            \draw[green,line width=1mm] (-2,-1) -- (0.5,1.5);
            \draw[black,line width=0.01mm] (-2,-1) -- (0.5,1.5);
            \draw [dashed, ->, green, thick] (0.5,1.5) -- (2.5,-0.5);
            
            \node[label=right:$\mathcal{J}_R^+$] (4) at (3.6,1.4) {};
            \node[label=right:$\mathcal{J}_R^-$] (5) at (3.6,-1.4) {};
            \node[label=below: $x_i^+$](6) at (1.5,-1.4) {};
            \node[label=below: $x^-$](6) at (-0.75,-2.2) {};
            \node[label=left:$\mathcal{J}_L^+$] (7) at (-3.6,1.7) {};
            \node[label=left:$\mathcal{J}_L^-$] (8) at (-3.6,-1.4) {};
             \node[label=below: $x_f^+$](9) at (2.1,-0.9) {};
             \node[label=below: $x^+$](10) at (0.75,-2.2) {};
             \node[label=above: $i_L^{+}$] at (-1.5,1.5) {};
            \node[label=above: $i_R^{+}$] at (1.5,1.5) {};
             \node[label=below: $i^{-}$] at (0,-3) {};
	      \node[label=below:$1$] at (-1.4,-0.4) {};
	      \node[label=below:$2$] at (0.2,-0.4) {};

            \draw[->,bend right=35] (4) to (2.3,0.8);
            \draw[->,bend left=35] (5) to (2.5,-1.0);
            \draw[->,bend left=35] (7) to (-2.3,0.8);
            \draw[->,bend right=35] (8) to (-2.5,-0.8);

        \end{tikzpicture}
    \end{center}
    \caption{Unavailability of Cauchy surface for left-moving modes}
    \label{fig:penrose05}
\end{figure}
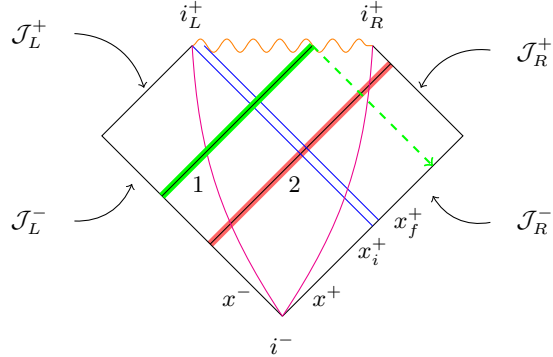

This effect can be more clearly understood using non-availability of Cauchy surfaces for future asymptotic observers. For that purpose let us consider a left moving timelike or null trajectory. Due to conformal flatness, a complete set of orthonormal mode functions on any surface orthogonal to them, can always be obtained as plane waves under a proper co-ordinatization (which extends as $(-\infty, \infty)$) of that surface. If we consider an orthogonal null surface for the left moving modes, before the formation of singularity  (the red surface labeled `2' in \ref{fig:penrose05}), the Bogoliubov transformation coefficients between that surface and $\mathcal{J}_{R}^{-}$  will be trivial, i.e., $(\alpha_{\Omega \omega} =\delta(\Omega -\omega), \beta_{\Omega \omega}=0)$.
However, once the singularity forms, any null surface relating a point on $\mathcal{J}_{L}^{-}$ to the singularity (e.g., green surface labeled `1' in \ref{fig:penrose05})), fails to causally connect with the entire spacetime. Also, any timelike observer has a domain of dependence, which is not the full spacetime. A portion of $\mathcal{J}_{R}^{-}$ (right to the dashed green line) is causally denied to the past of such surfaces. Therefore, the Bogoliubov coefficients assume a non-trivial form, such as in \ref{AlphaCGHS}, \ref{BetaCGHS} or \ref{AlphaCGHS-R-L}, \ref{BetaCGHS-R-L}, with a phase factor capturing the information about the region of the traced over modes. This is dynamically similar to a Rindler trajectory in a Minkowski spacetime, see \ref{fig:penrose06}, where an accelerated observer (thick green curve) reaches the future null infinity rather than future timelike infinity. For such observers too, the domain of dependence is only a part of the full spacetime. There is a horizon masking a portion 
of spacetime and hence the Bogoliubov transformation  coefficients between $\mathcal{J}_{R}^{-}$ and  $\mathcal{J}_{L'}^{+}$ (rather than the full $\mathcal{J}_{L}^{+} $ ) is non-trivial \cite{gravitation, Birrell:1982ix,Mukhanov-Winitzky} and the accelerated observers obtain a thermal spectrum for a vacuum state defined on  $\mathcal{J}_{R}^{-}$. However, there is no flux associated with this spectrum as the vacuum expectation value of the stress tensor vanishes identically  in the flat spacetime.

Similarly in the CGHS model, the region of spacetime $y^+ >y_i^+$ is dynamically made inaccessible to any left-moving time-like or null trajectories, which also is the case with such trajectories in the null collapse forming a Schwarzschild black hole \ref{fig:penrose02}.  However, in contrast to the Schwarzschild case, such observers in the CGHS  model, do not see any change of geometry hence they do not associate any mass to the ``black hole region'' as seen by them. The spacetime, they move in, throughout, is flat and such observers do not receive any flux of radiation, as the vacuum expectation value of the stress energy tensor which was vanishing on $\mathcal{J}_{R}^{-}$, stays put on zero, in the region $y^+ < y_i^+$.
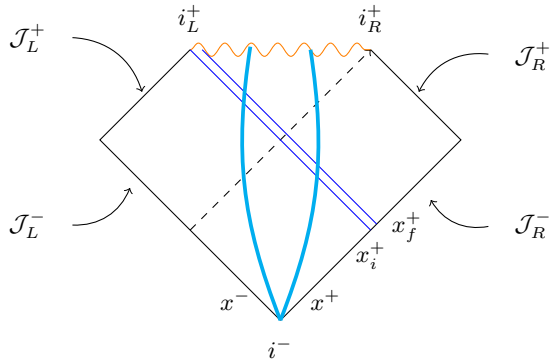
\begin{figure}[h]
    \begin{center}
        \begin{tikzpicture}[scale=0.8]
     
            \draw[dashed,->] (-1.5,-1.5) -- (0,0) -- (1.5,1.5);
        
            \draw[orange, snake it] (-1.5,1.5) -- (1.5,1.5);
              \draw[black] (0,-3) -- (3,0) -- (1.5,1.5);
            \draw[black] (0,-3) -- (-3,0) -- (-1.5,1.5);
            \draw[blue] (1.5,-1.5) -- (0,0) -- (-1.5,1.5); 
             \draw[blue] (1.60,-1.40) -- (-1.3,1.50);
             
            \draw[cyan,line width=0.5mm,bend left=15] (0,-3) to (-0.5,1.55);
            \draw[cyan,line width=0.5mm,bend right=15] (0,-3) to (0.5,1.5);

            
            \node[label=right:$\mathcal{J}_R^+$] (4) at (3.6,1.4) {};
            \node[label=right:$\mathcal{J}_R^-$] (5) at (3.6,-1.4) {};
            \node[label=below: $x_i^+$](6) at (1.5,-1.4) {};
            \node[label=below: $x^-$](6) at (-0.75,-2.2) {};
            \node[label=left:$\mathcal{J}_L^+$] (7) at (-3.6,1.7) {};
            \node[label=left:$\mathcal{J}_L^-$] (8) at (-3.6,-1.4) {};
             \node[label=below: $x_f^+$](9) at (2.1,-0.9) {};
             \node[label=below: $x^+$](10) at (0.75,-2.2) {};
               \node[label=above: $i_L^{+}$] at (-1.5,1.5) {};
            \node[label=above: $i_R^{+}$] at (1.5,1.5) {};
             \node[label=below: $i^{-}$] at (0,-3) {};

            \draw[->,bend right=35] (4) to (2.3,0.8);
            \draw[->,bend left=35] (5) to (2.5,-1.0);
            \draw[->,bend left=35] (7) to (-2.3,0.8);
            \draw[->,bend right=35] (8) to (-2.5,-0.8);

        \end{tikzpicture}
    \end{center}
    \caption{Freely falling observers in CGHS black hole spacetime}
    \label{fig:penrose07}
\end{figure}
As we discussed previously \ref{NumberOperator}, the right moving observers tuned to right moving modes, witness a thermal radiation flux at a temperature $1/\lambda$ which is independent of the matter content of the forming black hole and depends only on the other dimensionful parameter in the theory.

Following an identical path, the observers at $\mathcal{J}_{L}^{+}$ coupled to left-moving modes, observe a similar kind of Bogoliubov coefficients as their right-moving counterparts  \ref{AlphaCGHS}, \ref{BetaCGHS} but with the parameter exchange $|y_i^+| \leftrightarrow P^{+}/\lambda$ which mark the corresponding event horizons for such observers and appear as overall phases in the transformation coefficients. Hence, we see that such form of Bogoliubov coefficients \ref{AlphaCGHS-R-L}, \ref{BetaCGHS-R-L} are entirely due to tracing over of modes which lie in the causally denied region of spacetime, to ${\cal I}_L^+$ and not due to any geometry change. One can check that if the fraction of tracing over vanishes, which is marked by $|y_i^+|\rightarrow 0$, the Bogoliubov coefficients assume a trivial form, i.e., $\alpha _{\Omega \omega} \rightarrow \delta(\Omega-\omega)$ while $\beta _{\Omega \omega} \rightarrow 0$. Similarly, the Bogoliubov coefficients for the right-moving observers 
\ref{AlphaCGHS}, \ref{BetaCGHS}, assume a 
trivial form in the limit $P^+ \rightarrow 0$, marking the vanishing of event horizon, as well as, the amount of change of geometry suffered by such observers. Hence, for such observers effects of tracing over is indistinguishable from geometry change. Both these effects vanish simultaneously in that limit. Therefore, we see that the vacuum response for both left-moving or right-moving observers is indistinguishable from each other.  Late time environment for such observers on $\mathcal{J}_{L}^{+}$  or on $\mathcal{J}_{R}^{+}$ for vacuum state (of a test field)  is thermal. 

However, unlike the left-moving observers, the right moving observers also associate a flux with the radiation and hence the black hole region shrinks as a result of the evaporation. While left-moving observers do not associate any mass to the region beyond their horizon, the location of their horizon does not shrink and the ``black hole region'' does not evaporate for them. Classically speaking it creates a casual dilemma for the left moving inertial observers since they can not associate any source to the radiation they observe. Such observers are oblivious to an event horizon formed due to collapsing matter, from which they are unable to get any causal communication;  nor does the geometry reveal  such a development. However, quantum correlations of the global vacuum state seep through the horizon, which these observers are able to tap (reminiscent of what happens in the famous EPR paradox). It is  these correlations  which  the left-moving observers \emph{attribute, from their point-of-view,  as due to an excited state defined on the Hilbert space on their accessible region}.  Had there been no dynamics in the region $y^+ >y_i^+$, the full Cauchy surface would have become accessible to the left-moving observers (as well), making them standard inertial observers. Thus, the dynamics of the spacetime has generated an Unruh (like) effect for a set of inertial observers.

All the geodesic observers moving rightwards or remaining stationary at any finite value (with the exception of stationary observer at left infinity), on the other hand, end up in singularity and in this process have to undergo a geometry change (thick cyan curves in  \ref{fig:penrose07}), so they witness a combined effect and hence a flux of radiation, as in the Schwarzschild scenario \cite{Smerlak:2013sga,Singh:2014paa,Chakraborty:2015nwa}. Whereas the left-moving timelike observers end up at $i_L^+$, do not see any geometry change but encounter flux-free thermal atmosphere due to a dynamical emergence of a horizon, which washes out information of a section of initial field configuration, exactly in the spirit of the Rindler observer, as discussed previously.

Finally, we will comment on the back-reaction due to black hole evaporation and its implications for our result. To compute any kind of back-reaction in the spacetime in Einstein gravity, one needs to use an equation of the form $G_{ab}=\langle T_{ab}\rangle$. \textit{Neither side of this equation can be handled without additional assumptions and, obviously, the result will depend on the additional assumptions!} The left-hand-side identically vanishes in $D=2$ because Einstein tensor identically vanishes in two dimensions. The right hand side is: (i) divergent and needs to be regularized and (ii) depends on the quantum state which is chosen. The best we can do, therefore, is to give prescriptions to define both sides and the results will obviously \textit{depend on these prescriptions}. We discuss our set of choices while describing the pertinent issues in two-dimensional gravity below.

To get a non-zero left hand side, one postulates some non-Einsteinian form of the gravity action (like the CGHS action in \ref{CGHS_Action}) and vary the metric to get the equations of motion. Obviously, this is not $G_{ab}$ (which, of course, is zero) but can possibly act as a proxy for the same. We stress the fact that we are trying to model some gravitational features by some suitable dilatonic action and one cannot ignore the implicit ad-hocness in the procedure. For example, in such an approach, one also often takes the the normal ordered, non-covariant, expectation value  of the stress energy tensor as the classical source. However, in a two-dimensional spacetime, the expectation value of the stress energy tensor comes with a conformal anomaly term \cite{Fabbri:2005mw} as well. The anomaly term can then act as an extra source of stress energy and may lead to an evolution different from the classical case, if not accounted for properly. A prescription used in the literature is to add some extra terms (e.
g., Polyakov action, RST action terms) to the standard CGHS action corresponding to the conformal anomaly terms and obtain these {\it modified} equations of motion, as a result of the variational principle. But this new action will correspond to a different physical system, though one which can also be thought of as a proxy for gravity.  But in this prescription, flat spacetime will not be a solution to the vacuum state for the semiclassical equations due to the anomaly, which we consider somewhat unphysical. There is no unique way of handling this issue because, as we stressed before, $G_{ab}=0$ in $D=2$ and the left hand side which we work with to mimic $G_{ab}=\langle T_{ab} \rangle$ depends crucially on the model we use, with the hope that it can mimic aspects of $D=4$ gravity. We have chosen simple, physically well-motivated choices to do this as we describe next.

To analyze the right hand side, we need a scheme for defining a $c-$number stress tensor from the quantum operator and the vacuum state has to be motivated from some specific (geometric) considerations. We have \textit{defined} the vacuum state by a natural assumption: viz. that the geometry must remain flat when sourced by such a (vacuum) state. This is important because as we discussed, in two-dimensional spacetime, the expectation value of the stress energy tensor has a conformal anomaly term which can violate this criterion in general, if this term is not accounted for in the source terms in the right hand side. We will choose the state of the matter field such that one recovers the classical flat geometry prior to $x_i^+$, in order to remain true to the classical consideration. Therefore, we require $\exp(\rho)=\exp(\rho_{\textrm{flat}}) =1/\lambda^2x^+x^-$ in the region $x^+ <x_i^+ $. Thus, the classical values of the $T_{\pm \pm}$ are being realized by $ \langle T_{\pm \pm} \rangle$. 
We can further choose the boundary conditions for the set of initial states judiciously such that the contribution due to conformal anomaly exactly cancels out in the region of interest leading to a flat spacetime semi-classically, see for instance \cite{Vaz:1996kh, Ashtekar:2010hx, Hawking:1992cc}. For this judicious choice of family of quantum states \cite{Lochan:2016nbs}, the spacetime region $x^{+}<x_{i}^{+}$ remains unaffected by the conformal anomaly term. Therefore the results discussed in this work, use stability of the solution under these class of states. A more detailed analysis of conformal anomaly modified evaporating black holes can be found 
in \cite{Fabbri:2005mw}. Our scheme to deal with such terms will be reported in details elsewhere \cite{KLSCTP3}.
\section{FAQ: Demystifying some aspects of the analysis}

In this section we provide answers to some FAQ, which will help the reader to understand the content of this work from a better perspective. 

\begin{enumerate}
 \item {\it What does the non-trivial Bogoliubov coefficients for the left patch signify ?}\\
 
The Bogoliubov coefficients essentially signify the ``particle content'' of a field configuration as measured by suitable annihilation and creation operators between two sets of observers who are entitled to declare positive frequency excitations of the field. (In general, this ``particle content will be quite different from the particle content determined by detectors; for a discussion see e.g. \cite{Sriramkumar:1999nw}). If two set of observers are connected by trivial Bogoliubov coefficients, they must agree on the field content, and, in particular, the notion of the vacuum state. In the case we are interested in, the Bogoliubov coefficients are non-trivial for the left moving observer and hence (s)he will ascribe the `in' vacuum state of the quantum field for the spacetime to be filled by particles as determined by the expectation value of a number operator. Notably, such observers may well be left moving {\it inertial} observers.

\item {\it Will such observers, therefore find a swarm of particles hitting them ?}\\

The answer, as in the case of Unruh effect, is no. Since the left moving observer lives in a static patch of the spacetime, having time reversal symmetry, there is no outgoing flux for the relevant observers. These observers will find the state of the field to be just thermally populated. This is very much analogous to the experience of the Rindler observer, witnessing Unruh effect, but for the fact that the left moving observers in this CGHS spacetime are \textit{inertial} observers.

\item {\it There are other inertial observers too, such as asymptotic observers in Schwarzschild spacetime, $r=0$ observer in de Sitter. What is so special in this particular observation?}\\

Indeed there are other observers who do witness thermal radiations, in one way or another, despite being inertial. However, they are mostly in a spacetime with non-vanishing curvature. Any exploration of the spacetime region in the vicinity of their location will reveal the non-zero curvature to them. On the other hand, the observers discussed in this paper, live completely in a flat spacetime region throughout their lifetime. They do not have any spacetime curvature or acceleration to explain their experiences.

\item{\it The non-trivial Bogoliubov coefficients tell us that these observers declare the initial vacuum as excited state. This happens because of a null shell collapsing in their causal future. Is this not in violation of causality?}\\

The single most important thing to notice is that, the notion of Bogoliubov coefficients is global rather than local. In order to evaluate the Bogoliubov coefficients, one must have a knowledge regarding the full global geometry beforehand. So it is premature to contest causality from the non-triviality of the Bogoliubov coefficients. Causality in quantum field theory is protected by the commutation relations, which in this context, across the horizon, still vanishes. So based on that one can conclude that there is no breakdown of causality. The left moving observers just have access to (or get a signal from) the field configuration in a portion of the spacetime, which remains insufficient for them to conclude the field to be in the vacuum state. Just like based on non-trivial Bogoliubov coefficients of a Rindler observer, one can not object that the Unruh radiation lets the observer know what it will be doing in the future (i.e.,  observer will be eternally accelerating, making her incapable of changing her 
mind of whether to accelerate or not) in violation of causality, one can not object to causality in this set-up as well.

\item{\it Then how should one interpret these Bogoliubov coefficients?}\\

In the same spirit as the Bogoliubov coefficients in the case of Rindler observer. For example if we consider an accelerated observer in the right wedge of the Minkowski spacetime, that observer will be oblivious to the field configuration in the left wedge. Still if the field configuration is distorted only in the left wedge, it will carry imprints of that in the right wedge too, without breaking any causality. The crucial thing is that no information can propagate from left to the right wedge.

\item {\it How does the collapsing shell affect the left moving observers then?}\\

The collapsing shell ensures that the left moving observers do not have infinite proper time in the future. The global geometry is modified due the collapse of the null shell, making the left moving observers trajectory only finitely extendible in the future. Therefore, the mode-functions which provided a complete basis in past do not remain relevant in the future and one needs to find mode functions which provide a complete orthonormal basis for left moving observers in the future. Therefore the vacuum viewed in the light of new orthonormal basis appears as if it is excited. Again to stress the central point, the quantum correlations between events in the future and past of the shell remain non-zero, but their commutator vanishes, thus there is no causal violation. 

\item{\it What are the requirements which go into deriving the form of mode functions suited for future observers?}\\

For future observers, one needs to get a complete basis in the region accessible to them (just like Rindler observers, or for late time observers in black hole case for that matter). Further, we require a complete range of co-ordinates over which the variable characterizing the  plane wave runs, in order to get a complete basis (i.e., $e^{ikx}$ provides a complete basis \emph{only} if $x\in(-\infty, \infty)$. Given this fact, one can decompose the positive frequency modes in the future in terms of positive frequency and negative energy modes in the past to obtain the Bogoliubov coefficients between them. In the co-ordinate system which yield a complete basis, we also have a positive frequency mode-function in terms of co-ordinate time (just like e.g., early time mode function during inflation). Though this co-ordinate time is not a Killing one.

\item{\it What does the reduced density matrix for the left moving observer look like?}\\

Again, since the left moving observers are totally ignorant about the field configuration in the region which is future to the shell, they will have to trace over those configurations to obtain a reduced density matrix.  However, the Bogoliubov coefficients are of the form that they pronounce thermality only for high frequencies (like in Hawking radiation). For low frequencies, i.e at long wavelengths, one is able to see that the region explored by left moving observers is different from an eternally accelerated observer, leading to departure from thermality. The reduced density matrix is also expected to bear these features.

\item{\it Can such effects be present in other dimensions ?}\\
Our result depends essentially  on chirality of the dynamics in 2-dimensional dilaton gravity. In $D=2$ we have a solution in which observers moving in one direction do not feel the presence of black hole which has formed as a result of a collapse. Currently, we do not know of any other solutions in any theory of gravity  which has such a feature. Evidently, due to the existence of standard no-hair theorems, Einstein gravity will not lead to such an asymmetric black hole configuration. However, whether  any other theory of gravity can lead to such a solution in higher dimensions, is a question which we hope to address in a future study.

\end{enumerate}
\section{Concluding Remarks}

In this work, we have shown that pure quantum correlations can play the role of a source of a thermal bath in a flat region of spacetime. There is no classical source otherwise, to describe the thermal ambiance of the quantum field, which a section of inertial observers in the spacetime and into. The inertial observers in the flat region are unaware of an event horizon formation due to the collapsing matter, which lies to their causal future. Therefore, unlike their right moving counterparts, they are unable to witness any black hole formation, and do not have a explanation to their thermal environment other than describing the state to be a non-vacuum one. It is well known that, quantum correlations (e.g., of the global vacuum state) do not vanish beyond the light cone and it is these fluctuations,  which these 
observers are actually detecting. It is the partially traced over quantum correlations (due to presence of an horizon the existence of which these observers do not know), which the left moving observers \emph{falsely} associate with an excited state, defined on the Hilbert space on their accessible region, since clearly the quantum fluctuations of the vacuum state defined on their Hilbert space would have had different signatures. 

This ``thermality she gets is not through any geometry change, or through any artifact of her non-inertial motion \emph{but through quantum correlations} of the ambient field. However by no means is this a case of causality violation any more than there is causality violation in the EPR situation, with the counterpart of the thermal excitations lying beyond the horizon. Only a pair of observers across the horizon will have the complete information about the vacuum structure, not any one of them individually. This is, of course, in stark contrast to the right moving observers, who, knowing the field equations of dilaton gravity, will be able to associate a change of geometry and hence interpret it as the source, essentially as the phenomenon of black hole evaporation with no apparent surprise.

Thus, we have shown quantum correlations as a probe of maximal extendibility of the spacetime. Any probe of such quantum correlations will reveal the differences between the vacuum of a maximally extended and maximally extensible spacetime, which remain hidden classically. In addition to these correlations, if some {\it classical channels} \cite{Henderson, Costa2016} are also present, we can in principle utilize them to probe the geometry of causally inaccessible regions more effectively in higher dimensions as well. We will pursue these issues in a subsequent work. 
\section{Acknowledgements}
The authors thank Alessandro Fabbri for useful comments on the preprint version of the paper. Research of KL is supported by
INSPIRE Faculty Fellowship grant by Department of Science and Technology, Government of India. Research of SC is funded by a SERB-NPDF grant (PDF/2016/001589) from SERB, Government of India. The research of TP is partially supported by the J.C. Bose research grant of the Department of Science and Technology, Government of India. 
\bibliography{Gravity_1_full,Gravity_2_partial,My_References}

\bibliographystyle{./utphys1}
\end{document}